\documentclass[letterpaper,twocolumn,10pt]{article}
\pdfoutput=1
\usepackage{usenix,epsfig,endnotes}
\usepackage[english]{babel}
\usepackage{graphicx}
\usepackage{balance}
\usepackage{times}
\usepackage{algorithm}
\usepackage[noend]{algorithmic}
\usepackage{subfigure}
\usepackage{comment,amsthm,amssymb}
\usepackage{xspace,url}

\newcommand{\oursys}{{\em G-thinker}\xspace} 
\newcommand{\prob}{subgraph finding\xspace} 

\begin{document}

\date{}

\title{\Large \bf G-thinker: Big Graph Mining Made Easier and Faster}

\author{
{\rm Da Yan$^{\S\dag1}$
, Hongzhi Chen$^{\S2}$, James Cheng$^{\S3}$, M.\ Tamer \"{O}zsu$^{\ddag4}$, Qizhen Zhang$^{\S5}$, John C.\ S.\ Lui$^{\S6}$}
\vspace{1.6mm}\\
\fontsize{10}{10}\selectfont\itshape\rmfamily $^\S$Department of Computer Science and Engineering, The Chinese University of Hong Kong\\
\fontsize{9}{9}\selectfont\ttfamily\upshape \{$^1$yanda, $^2$hzchen, $^3$jcheng, $^5$qzzhang, $^6$cslui\}@cse.cuhk.edu.hk\\
\fontsize{10}{10}\selectfont\itshape\rmfamily $^\dag$Department of Computer and Information Sciences, The University of Alabama at Birmingham\\
\fontsize{9}{9}\selectfont\ttfamily\upshape $^1$yanda@uab.edu\\
\fontsize{10}{10}\selectfont\itshape\rmfamily $^\ddag$David R.\ Cheriton School of Computer Science, University of Waterloo\\
\fontsize{9}{9}\selectfont\ttfamily\upshape $^4$tozsu@uwaterloo.ca
}

\maketitle

\thispagestyle{empty}

\subsection*{Abstract}
This paper proposes a general system for computation-intensive graph mining tasks that find from a big graph all subgraphs that satisfy certain requirements (e.g., graph matching and community detection). Due to the broad range of applications of such tasks, many single-threaded algorithms have been proposed. However, graphs such as online social networks and knowledge graphs often have billions of vertices and edges, which requires distributed processing in order to scale. Unfortunately, existing distributed graph processing systems such as Pregel and GraphLab are designed for data-intensive analytics, and are inefficient for computation-intensive graph mining tasks since computation over any data is coupled with the data's access that involves network transmission. We propose a distributed graph mining framework, called \oursys, which is designed for computation-intensive graph mining workloads. \oursys provides an intuitive graph-exploration API for the convenient implementation of various graph mining algorithms, and the runtime engine provides efficient execution with bounded memory consumption, light network communication, and parallelism between computation and communication. Extensive experiments were conducted, which demonstrate that \oursys is orders of magnitude faster than existing solution, and can scale to graphs that are two orders of magnitude larger given the same hardware resources.

\section{Introduction}\label{sec:intro}
We focus on a class of graph mining problems, namely \prob problems, which aim to find all subgraphs in a graph that satisfy certain requirements. It may enumerate (or count) all of the subgraphs, or find only those subgraphs with top-$k$ highest scores, or simply output the largest subgraph. Examples of \prob problems include \emph{graph matching}~\cite{match3}, \emph{maximum clique finding}~\cite{clique03}, \emph{maximal clique enumeration}~\cite{clique73}, \emph{quasi-clique enumeration}~\cite{quasiclique}, \emph{triangle listing and counting}~\cite{yufei_triangle}, and \emph{densest subgraph finding}~\cite{densegraph}. These problems have a wide range of applications including social network analysis~\cite{PattilloYB13eor,socialnet}, searching knowledge bases~\cite{KasneciSIRW08icde,WuLWZ12sigmod} and biological network investigation~\cite{HeS08sigmod,ZouCO09pvldb}. Although many serial algorithms have been proposed to solve these problems, they cannot scale to big graphs such as social networks and knowledge graphs. Moreover, it is non-trivial to extend these algorithms for parallel processing, because (1)~the input graph itself may be too big to fit in the memory of one machine, (2)~a serial algorithm checks subgraphs using backtracking, where only one candidate subgraph is constructed (incrementally from the previous candidate subgraph) and examined at a time; a parallel algorithm that checks many subgraphs simultaneously in memory many cause memory overflow. The issue is further complicated in the distributed setting due to the high cost of accessing large amounts of remote data.

A plethora of distributed systems have been developed recently for processing big graphs~\cite{dbs}, but most of them adopt a think-like-a-vertex style (or, vertex-centric) computation model~\cite{giraph,graphx,gps,da_www} which is pioneered by Google's Pregel~\cite{pregel}. These systems only require a programmer to specify the behavior of one generic vertex (e.g., sending messages to other vertices) when developing distributed graph algorithms, but the resulting programs are usually data-intensive. Specifically, the processing of each vertex is triggered by incoming messages sent (mostly) from other machines, and the CPU cost of vertex processing is negligible compared with the communication cost of message transmission. Moreover, subgraph finding problems operate on subgraphs rather than individual vertices, and it is unnatural to translate a subgraph finding problem into a vertex-centric program.

As a result, big graph analytics research mostly focuses on problems that naturally have a vertex-program implementation, such as PageRank and shortest path, and few work has been devoted to large-scale subgraph finding. To our knowledge, only two existing systems, NScale~\cite{nscale} and Arabesque~\cite{arab}, attempted to attack large-scale subgraph finding with a subgraph-centric programming model. Unfortunately, their execution engines still mine subgraphs in a data-intensive manner. Specifically, they construct all candidate subgraphs in a synchronous manner (i.e., large subgraphs are constructed from small ones), before the actual computation that examines these subgraphs. Materializing candidate subgraphs incurs large network and storage overhead (while CPU is under-utilized), and the actual computation-intensive mining process is delayed until the costly subgraph materialization is completed.

Another critical drawback of existing distributed frameworks is that they process each individual subgraph as an independent task, which loses many optimization opportunities. For example, if multiple subgraphs on a machine contain a common vertex $v$, their tasks may share $v$'s information (e.g., adjacent edges). But in existing systems, each subgraph will maintain its own copy of $v$'s information (likely received from another machine), leading to redundant communication and storage. The synchronous execution model of existing systems is also prone to the straggler problem, due to imbalanced workload distribution among different machines.

In this paper, we identify the following five requirements that a distributed system for \prob should satisfy in order to be efficient and user-friendly:
\begin{itemize}
\item The programming interface should be {\bf subgraph-centric}.
\item {\bf Computation-intensive} processing should be native to the programming model. For example, a programmer should be able to backtrack a portion of graph to examine candidate subgraphs like in a serial algorithm, without materializing and transmitting any subgraph.
\item There should exist {\bf no global synchronization} among the machines, i.e., the processing of different portions of a graph should not block each other.
\item Since a \prob algorithm checks many (possibly overlapping) subgraphs whose cumulative volume can be much larger than the input graph itself, it is important to schedule the subgraph-tasks properly to keep the {\bf memory usage bounded} at any point of time.

    Obviously, each machine should stream and process its subgraphs on its local disk (if memory is not sufficient), to minimize network and disk IO overhead.
\item Subgraphs on a machine that contain a common vertex $v$ should be able to share $v$'s information (e.g., adjacent edges), to {\bf avoid redundant data transmission and storage}.
\end{itemize}

Based on these criteria, we designed a novel subgraph-centric system, called \oursys, as a unified framework for developing scalable algorithms for various \prob problems. To write a \oursys program, a user only needs to specify how to grow a portion of the input graph $g$ by pulling $g$'s surrounding vertices, and how to process $g$ (e.g., by backtracking). Communication and execution details in \oursys are transparent to end users. In \oursys, each machine only keeps and processes a small batch of tasks in memory at any time (to achieve high throughput through task batching while keeping memory usage bounded). Subgraphs that are waiting to be processed (e.g., to grow its frontier by pulling remote vertices) are buffered in a disk-based priority queue. The priority queue is organized by a min-hashing based task scheduling strategy, in order to maximize the opportunity that the processing of different subgraphs share common vertices (including their adjacent edges) that are cached in the local machine.

We have used \oursys to develop significantly more efficient and scalable solutions for a number of \prob problems, including triangle counting, maximum clique finding, and graph matching. Compared with existing systems, \oursys is up to hundreds of times faster and scales to graphs that are two orders of magnitude larger given the same hardware resources.

The rest of this paper is organized as follows. We motivate the need of a computation-intensive subgraph finding framework with the problem of maximal clique enumeration in Section~\ref{sec:example}, and then explain why existing systems are inefficient in Section~\ref{sec:related}. Section~\ref{sec:overview} provides an overview of the design of \oursys, Section~\ref{sec:api} introduces the programming interface of \oursys, and Section~\ref{sec:app} illustrates how to write application programs in \oursys. We present the implementation of \oursys in Section~\ref{sec:system}, and report experimental results in Section~\ref{sec:results}. Finally, we conclude the paper in Section~\ref{sec:conclude}.

\section{A Motivating Example}\label{sec:example}
A common feature of \prob problems is that, the computation over a graph $G$ can be decomposed into that over subgraphs of $G$ that are often much smaller (called {\bf decomposed subgraphs}), such that each result subgraph is found in exactly one decomposed subgraph. In other words, the decomposed subgraphs partition the search space and there is no redundant computation. We illustrate by considering maximal clique enumeration, which serves as our running example. Table~\ref{gnote} summarizes the notations used throughout this paper.

\begin{table}[t]
\caption{Notation Table}\label{gnote}
\vspace{1mm}
\centering
\includegraphics[width=0.9\columnwidth]{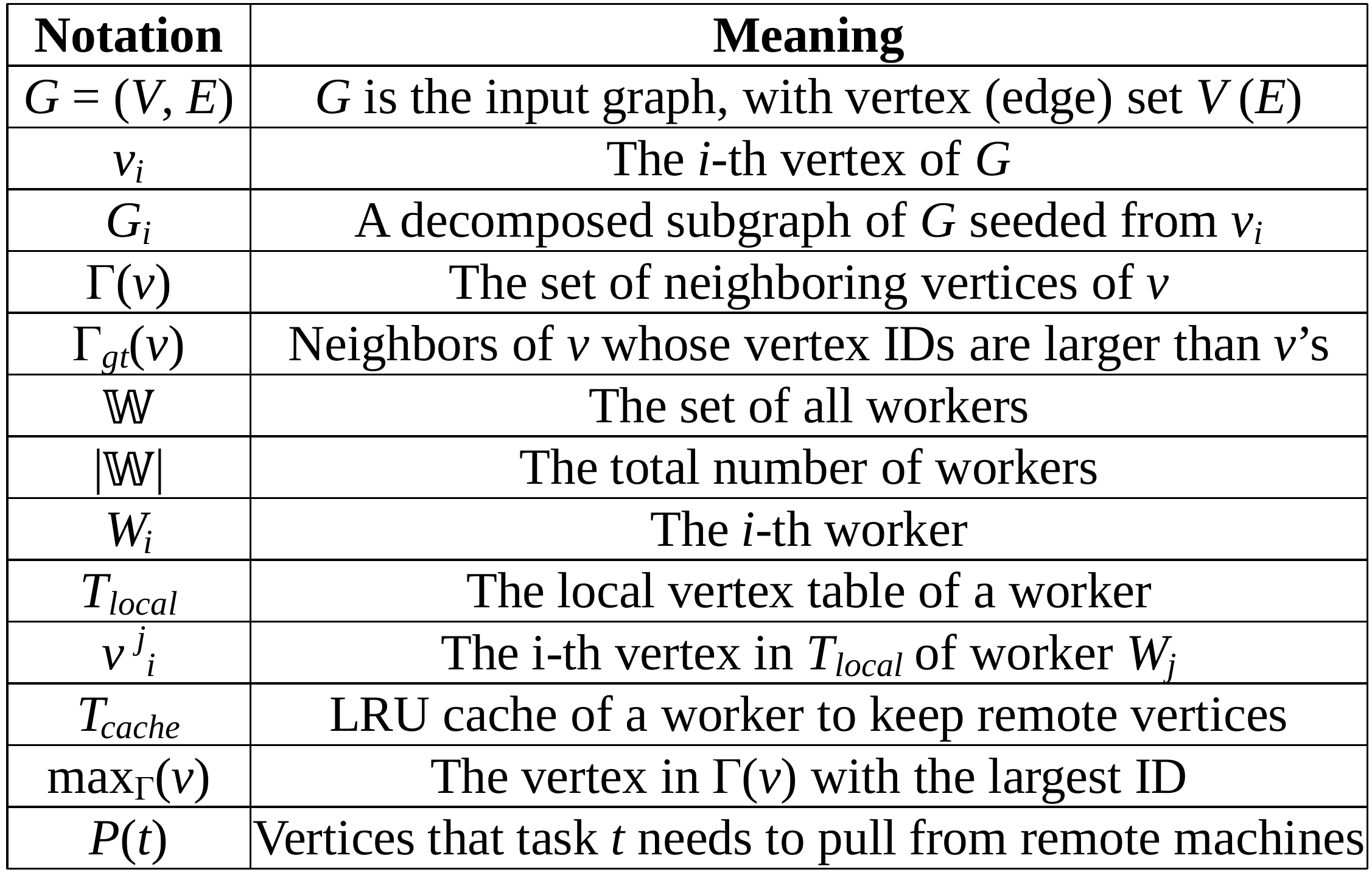}
\vspace{-4mm}
\end{table}

\vspace{1mm}

\noindent{\bf Example: Maximal Clique Enumeration.} We decompose a graph $G=(V, E)$ into a set of $G$'s subgraphs $\{G_1, G_2, \ldots, G_n\}$, where $G_i$ is constructed by expanding from a vertex $v_i\in V$. Let us denote the neighbors of a vertex $v$ by $\Gamma(v)$. If we construct $G_i$ as the subgraph induced by $\{v_i\}\cup\Gamma(v_i)$ ($G_i$ is called $v_i$'s 1-ego network), then we can find all cliques from these 1-ego networks since any two vertices in a clique must be neighbors of each other. However, a clique could be double-counted.

Let us define $\Gamma_{gt}(v)=\{u\in\Gamma(v)\,|\,u>v$\} where vertices are compared according to their IDs. To avoid redundant computation, we redefine $G_i$ as induced by $\{v_i\}\cup\Gamma_{gt}(v_i)$, i.e., $G_i$ does not contain any neighbor $v_j<v_i$. This is because any clique containing both $v_i$ and $v_j$ has already been computed when processing $G_j$. Obviously, any clique $C$ (let the smallest vertex in $C$ be $v_i$) is only computed once, i.e., when $G_i$ is processed. \qed

\vspace{1mm}

We can distribute these decomposed subgraphs to different machines, so that each decomposed subgraph is processed using a serial backtracking algorithm to find cliques without network communication. Since the computation complexity of maximal clique enumeration is exponential to graph size, the computation cost of processing $G_i$ is super-linear (to $G_i$'s size) with a small constant (i.e., computation-intensive), while the transmission cost of creating $G_i$ is linear with a large constant (due to limited network transmission rate). Thus, the computation cost and communication cost strike a balance when $G_i$ is sufficiently large, and overlapping computation and communication over decomposed subgraphs significantly improves the overall performance.

However, since real graphs often follow power-law degree distribution, there may exist some vertex $v_i$ with a very high degree, thus generating a large $G_i$. Due to high computational complexity, the machine processing $G_i$ may becomes the straggler that keeps processing $G_i$ while other machines finish their tasks and become idle. To tackle this problem, a system should allow $G_i$ to be further decomposed, so that the resulting decomposed subgraphs can be distributed to different machines for processing. In maximal clique enumeration, we can decompose $G_i$ exactly as how we decompose $G$, conditioned on that $v_i$ is already in any clique found therein. The decomposition may recurse by looking at more vertices until the resulting subgraphs are small enough for balanced workload distribution.

As we shall see in Section~\ref{sec:app}, the above ideas generalize to numerous \prob problems.

\section{Limitations of Related Work}\label{sec:related}
In this section, we review existing distributed solutions to \prob, and explain their weaknesses.

\vspace{1mm}

\noindent{\bf Vertex-Centric In-Memory Solutions.} Most vertex-centric systems are in-memory systems, where vertices (along with their adjacency lists) are partitioned among different machines in a cluster and kept in memory~\cite{giraph,graphlab,powergraph,graphx,gps,da_www}. Vertices communicate with each other by message passing, and messages are also buffered in memory to avoid slow disk access.

However, the vertex-centric API is not suitable for \prob, and each vertex $v_i$ needs to communicate with its surrounding vertices in a breadth-first manner (one more hop per iteration) to get their information for constructing $G_i$. The solution cannot scale to large graphs since the total volume of (possibly overlapping) decomposed subgraphs may easily exceed the memory capacity of a cluster. Vertex-centric systems also do not provide any mechanism for decomposed subgraphs to share common vertex's information\footnote{Pregel's message combiner~\cite{pregel} does not help, since it is to aggregate messages towards the same target vertex, while we consider getting information from the same source vertex.}.

The key problem is, nevertheless, that vertex-centric computation is mainly for data-intensive computation, and generates a large number of messages to transmit for \prob (e.g., for pulling vertices to construct decomposed subgraphs). We call the problem as {\bf communication-in-the-chain}. In fact, \cite{ppa} indicates that a vertex-centric program is most scalable if each iteration requires linear computation and communication cost, and it runs for a small number of iterations. This essentially implies that vertex-centric systems are for graph problems with a low computational complexity.

\vspace{1mm}

\noindent{\bf Vertex-Centric Disk-Based Solutions.} The prohibitive memory requirement can be eliminated using a disk-based system. For example, MapReduce~\cite{mapred} can be used to simulate vertex-centric graph computation (e.g., message sending \& receiving)~\cite{mapredPregel}, and Pregelix~\cite{pregelix} translates a vertex-centric program into a dataflow execution plan for out-of-memory processing. However, the large amount of intermediate data (including messages and subgraphs) need to be dumped to disk and then loaded back for each iteration of synchronous computation, making the running time prohibitive. We call the problem as {\bf disk-in-the-chain}, which adds upon the communication-in-the-chain problem already suffered by a vertex-centric model. In fact, MapReduce even writes intermediate data to Hadoop Distributed File System (HDFS), which is much slower than local disk writes since HDFS replicates each data block on three machines for fault tolerance (termed the {\bf remote write problem}).

These synchronous frameworks also prevent the computation-intensive processing of decomposed subgraphs from beginning until all decomposed subgraphs are synchronously constructed, leading to CPU under-utilization. Data sharing among subgraphs is also not possible since a subgraph is processed by a reducer.

\vspace{1mm}

\noindent{\bf Systems with Subgraph-Based API.} Recently, NScale~\cite{nscale} and Arabesque~\cite{arab} attempted to attack subgraph finding problems through a subgraph-based API rather than a vertex-centric one. Albeit becoming more user-friendly, the execution engines of these systems still perform data-intensive processing like vertex-centric solutions mentioned before, and they actually introduce new performance issues.

NScale~\cite{nscale} uses the MapReduce solution we mentioned above, and it brings additional overheads. NScale only supports the top-level decomposed subgraphs, and there is no mechanism to balance workload through recursive decomposition. Assuming that each $G_i$ spans the $k$-hop neighborhood around $v_i$, then NScale first constructs all decomposed subgraphs using $k$ rounds of MapReduce. The large number of decomposed subgraphs are then packed into larger compact subgraphs, each of which can fit in the memory of a reducer. Vertices common to multiple decomposed subgraphs are stored only once in their packed subgraph. Finally, each compact subgraph is distributed to a reducer, which processes all decomposed subgraphs packed in the compact subgraph in memory. Obviously, NScale suffers from all the performance issues of a MapReduce-based vertex-centric solution; moreover, NScale further packs decomposed subgraphs through expensive disk-based computation, and it is very likely that the cost of packing $G_i$ already surpasses that of processing $G_i$ right after it is constructed in memory.

Arabesque~\cite{arab} proposed an embedding-centric model where an embedding is a subgraph of the input graph $G$. Arabesque requires the entire $G$ to reside in the memory of every machine, and constructs and processes subgraphs iteratively. In the $i$-th iteration, it grows the set of embeddings with $i$ edges/vertices by one adjacent edge/vertex, to construct embeddings with $(i+1)$ edges/vertices for processing. New embeddings that pass a filtering condition are further processed and then passed to the next iteration. For example, to find cliques, the filtering condition checks whether an embedding $e$ is a clique; if so, $e$ is reported and passed to the next iteration to grow larger clique candidates.

Unfortunately, Arabesque suffers from new performance and scalability issues. Firstly, while previous solutions still permit efficient backtracking within each decomposed subgraph, Arabesque materializes and transmits every single candidate subgraph it examines. Arabesque also compresses/decompresses the large number of materialized embeddings using a data structure called ODAG to save space, which consumes additional CPU cycles. To additionally support frequent subgraph pattern mining, automorphism checking is performed for every newly-expanded embedding to avoid generating duplicate embeddings, which adds unnecessary overhead for \prob. Finally, since $G$ resides in the memory of every machine, scalability is limited by the memory space of a single machine.

\vspace{1mm}

\noindent{\bf Other Systems.} Blogel~\cite{blogel} and Giraph++~\cite{giraph++} proposed a block-centric model which partitions a graph into {\bf disjoint} subgraphs called blocks to be distributed among machines for iterative processing, eliminating the need of communication inside each block. However, these systems do not target \prob problems, but rather the acceleration of vertex-centric models.

\section{System Overview}\label{sec:overview}
We now overview the design of \oursys, including its programmming model and system components.

\vspace{1mm}

\noindent{\bf Programming Model.} \oursys performs computation on subgraphs. Each subgraph $g$ is associated with a {\bf\em task}, which performs computation on $g$ and grows $g$ when needed. \oursys grows subgraphs starting from a set of seed vertices in $V$. For example, in clique enumeration, one may create a task from each vertex $v_i\in V$, which forms the initial subgraph $g$ containing only $v_i$; the task grows $g$ into  $G_i$ by {\bf\em pulling} vertices in $\Gamma_{gt}(v_i)$ along with their adjacent edges, and then enumerates cliques in $G_i$. In case $G_i$ is too big, users may instead further decompose $G_i$ and create new tasks associated with the newly decomposed subgraphs, which can then be distributed to different machines to improve load balancing.

\vspace{1mm}

\noindent{\bf \oursys Components.} A \oursys program runs on a cluster of workers, $\mathbb{W}=\{W_1, W_2, \ldots\}$, where each worker is a basic computing unit that processes its assigned tasks in serial, and a machine may run multiple workers. Each worker alternates between subgraph-centric task computation and vertex pulling (into subgraphs), both are processed in batches. To be memory efficient, we keep the memory requirement of each worker at approximately $O(d_{avg}\cdot\frac{|V|}{|\mathbb{W}|})$, where $d_{avg}$ is the average vertex degree.

\begin{figure}[t]
    \centering
    \includegraphics[width=\columnwidth]{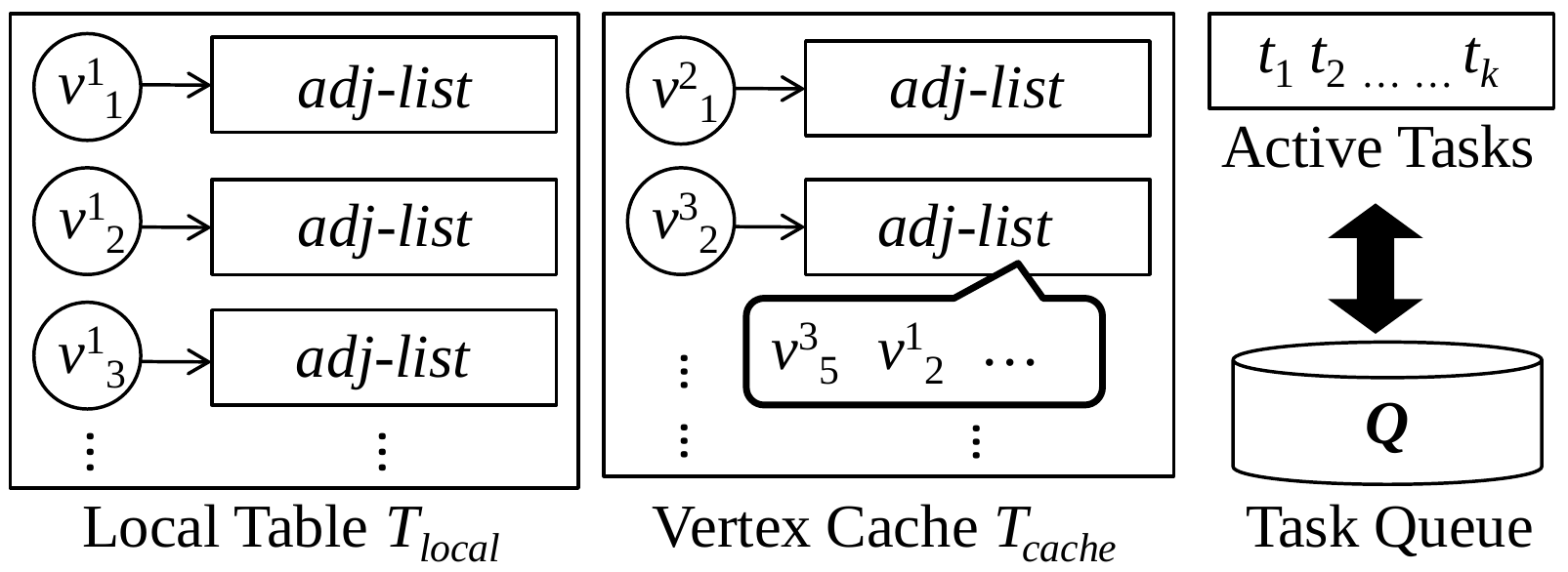}
    \caption{Components of Worker $W_1$}\label{worker1}
    \vspace{-4mm}
\end{figure}

\oursys partitions the vertices in $V$ (along with their adjacency lists) among different workers, and each worker maintains its assigned vertices in a local table $T_{local}$. Let us denote the $i$-th vertex maintained in $T_{local}$ of worker $W_j$ by $v^j_i$. Figure~\ref{worker1} shows the components of $W_1$, where we can see that $T_{local}$ maintains vertices $v^1_1, v^1_2, v^1_3, \ldots$; each vertex also keeps its neighbors in its adjacency list, so that it can pull its neighbors (along with their adjacency lists) from $T_{local}$ of other workers, by providing the neighbor IDs. The local tables of all workers collectively constitute a distributed key-value store where key is the ID of a vertex $v$ and value is $v$'s adjacency list $\Gamma(v)$.

When a vertex is pulled (along with its adjacency list) from another worker, it is not directly added to the subgraph of the requesting task; instead, it is put in an \emph{LRU cache} $T_{cache}$. The cache keeps the non-local vertices (i.e., not in $T_{local}$) that are previously received, so that a non-local vertex can be shared by all the tasks that pull it. It is up to the user to decide whether the task's subgraph should be updated, and if so, what information of that vertex should be added to the subgraph.

As Figure~\ref{worker1} shows, $W_1$'s $T_{cache}$ keeps non-local vertices $v^2_1, v^3_2, \ldots$ (along with their adjacency lists), which are pulled by local tasks previously executed at $W_1$. Note that the adjacency list of a non-local vertex may contain a local vertex, such as $v^1_2$ in the adjacency list of vertex $v^3_2$ in $T_{cache}$ in Figure~\ref{worker1}.

During subgraph-centric computation, when a task requires the adjacency list of a vertex $u$, if $u$ is in $T_{local}$ or $T_{cache}$, the task can directly obtain $\Gamma(u)$ by table lookup. Otherwise, the task needs to first pull $u$ from $T_{local}$ of $u$'s worker into the cache table $T_{cache}$, before accessing it.

As Figure~\ref{worker1} shows, each worker also maintains an \emph{in-memory task buffer} for keeping tasks that are currently being processed, and a \emph{disk-based task queue} for keeping tasks that are waiting to be processed. This design allows tasks to be processed with high throughput and less redundant communication, as we shall discuss next.

\vspace{1mm}

\noindent{\bf Batch Processing \& Communication Reduction.} A task usually only generates a small number of pull-requests at a time, and sending small messages wastes network bandwidth. Therefore, in \oursys, a worker fetches a batch of tasks from the disk-based task queue at each time, sends their pull-requests together, receives all the requested vertices, and then processes these tasks. Tasks that need to pull more vertices are then added to the task queue for further processing.

Batch processing hides the round-trip delay of each task's pull-requests, since if tasks are processed one at a time, each task needs to wait for its requested vertices to arrive, which wastes CPU cycles. Batch processing also reduces redundancy in communication. Specifically, if many tasks in a batch pull a remote vertex $u$, only one pull-request needs to be sent, and $\langle u, \Gamma(u)\rangle$ will be received only once and cached in $T_{cache}$ for access by all these tasks. This is in contrast to existing solutions like NScale, where $\langle u, \Gamma(u)\rangle$ needs to be transmitted to every subgraph $g$ that requires it. We organize the task queue $Q$ using locality sensitive hashing (detailed in Section~\ref{sec:system}), to increase the probability that tasks fetched from $Q$ share common vertices to pull.

To further reduce communication, \oursys allows a user to prune useless items in $\Gamma(v)$ before responding $\langle v, \Gamma(v)\rangle$ to a worker that pulls $v$. For example, in clique enumeration, a vertex $v$ only needs to respond to a pull-request with $\Gamma_{gt}(v)$ instead of the entire $\Gamma(v)$.

\vspace{1mm}

\noindent{\bf Memory Cost Analysis.} Since \oursys partitions the vertices evenly among the workers, $T_{local}$ of each worker contains around $O(|V|/|\mathbb{W}|)$ vertices. To keep the memory consumption bounded by $O(d_{avg}\cdot|V|/|\mathbb{W}|)$, we also set the capacity of $T_{cache}$ to cache at most $O(|V|/|\mathbb{W}|)$ vertices at any time, and vertex eviction is based on the LRU (Least Recently Used) policy. As for the memory requirement of tasks, since each task keeps a subgraph $g$, one cannot afford to keep all tasks in memory (e.g., consider all maximal cliques of a graph). Our solution is to keep only a small number (e.g., 1000) of active tasks in memory for batch processing, so that their small subgraphs consume $O(d_{avg}\cdot|V|/|\mathbb{W}|)$ memory space.

\vspace{1mm}

\noindent{\bf Computation \& Communication Cost Analysis.} Each task in \oursys (1)~pulls required vertices to its subgraph (linear communication cost, and pull requests can further be shared with other tasks) and (2)~then performs higher-complexity computation on the subgraph in local machine. Step~(2) is computation-intensive and avoids any communication when exploring the large search space by backtracking.

To overlap communication (i.e., Step~(1)) with computation (i.e., Step~(2)), \oursys treats tasks independently. Different tasks can have different progress, and no synchronization among all machines is required. \oursys proceeds the computation of a task as long as all its requested vertices are locally accessible, and the only communication type in \oursys is point-to-point communication between two workers for vertex pulling, and dynamic task (or decomposed subgraph) reassignment if load balancing is enabled.

\section{Programming Interface}\label{sec:api}
\begin{figure}[t]
    \centering
    \includegraphics[width=\columnwidth]{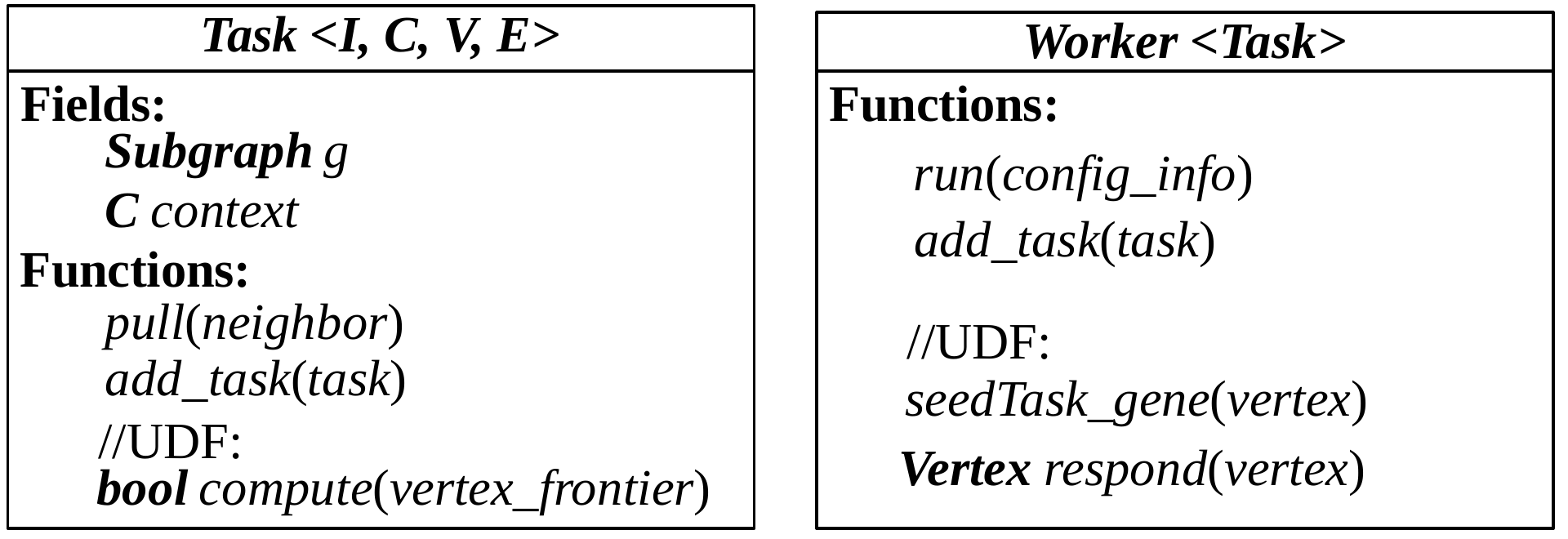}
    \caption{Programming Interface of \oursys}\label{api_simple}
    \vspace{-4mm}
\end{figure}

\oursys is written in C++, and it defines two important base classes, {\em Task} and {\em Worker}, as sketched in Figure~\ref{api_simple}. To write a \oursys program, a user needs to subclass {\em Task} and {\em Worker} with their template arguments properly specified, and implement their abstract functions according to the application logic; these functions are called \emph{user-defined functions} (UDFs). We remark that although we use C++ terminology here such as ``template'', the API is general enough to be implemented in any object-oriented language (e.g., ``generic types'' in Java).

\vspace{1mm}

\noindent{\bf Data Types.} As Figure~\ref{api_simple} shows, the {\em Task} class takes four template arguments $<$$I$$>$, $<$$C$$>$, $<$$V$$>$ and $<$$E$$>$. Among them, $<$$I$$>$, $<$$V$$>$ and $<$$E$$>$ specify the data types of vertices and edges: (1)~$<$$I$$>$: the type of vertex ID; (2)~$<$$V$$>$: the type of vertex attribute; (3)~$<$$E$$>$: the type of the attribute of an adjacency list item. Other system-defined types (e.g., those for subgraph, vertex, and adjacency list) are automatically derived by {\em G-thinker} from them, and can be directly used in the UDFs once a user specifies these three template arguments.

\begin{figure}[t]
    \centering
    \includegraphics[width=0.78\columnwidth]{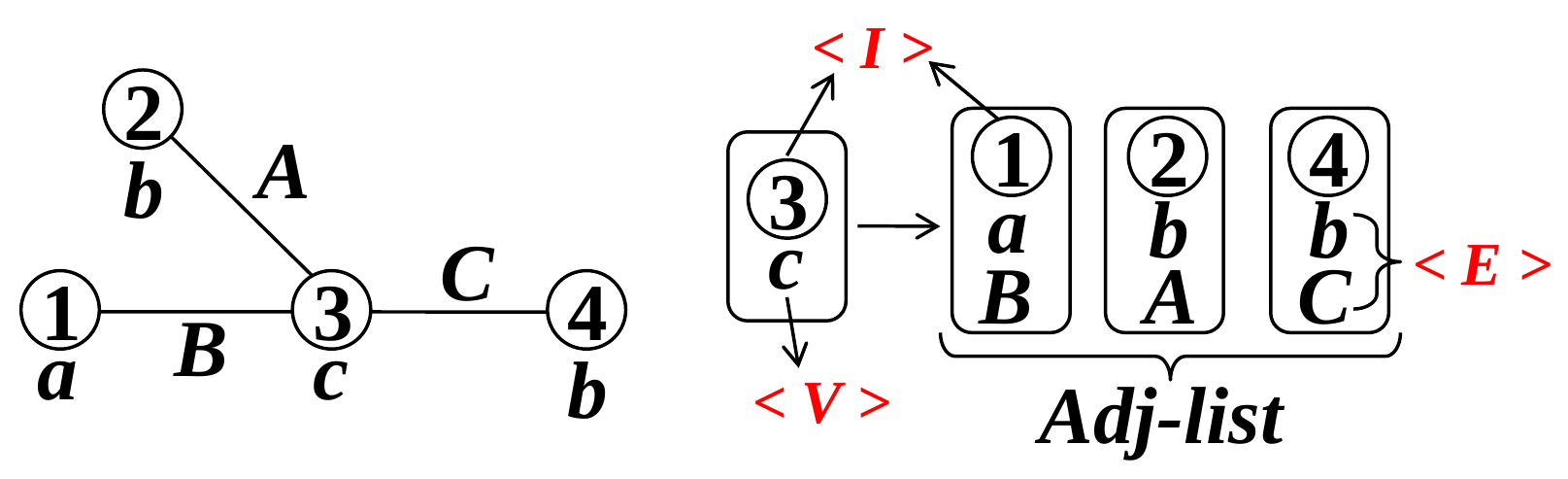}
    \caption{Data Types in \oursys}\label{api_example}
    \vspace{-4mm}
\end{figure}

Figure~\ref{api_example} illustrates the inferred system-defined types. Specifically, a subgraph is shown on the left, which consists of a table of vertices (stored with their adjacency lists). The structure of Vertex~3 is shown on the right, where the vertex is stored with its ID (of type $<$$I$$>$) and a vertex label ``c'' (of type $<$$V$$>$), and an adjacency list. Each item in the adjacency list is stored with a neighbor ID (of type $<$$I$$>$) and an attribute (of type $<$$E$$>$) indicating the label of the neighbor and the edge label. For example, the first item corresponds to Vertex~1 with label ``a'', and the edge label of $(3, 1)$ is ``B''. Attributes (i.e., $<$$V$$>$ and $<$$E$$>$) are optional and are not needed for finding subgraphs with only topology constraints (e.g., triangles, cliques, and quasi-cliques).

\vspace{1mm}

\noindent{\bf The Task class.} The {\em Task} class has another template argument $<$$C$$>$ that specifies the type of context information for a task $t$, which can be, for example, $t$'s iteration number (a task in \oursys proceeds its computation in iterations). Each {\em Task} object $t$ maintains a subgraph $g$ and the user-specified {\em context} object (of type $<$$C$$>$). The {\em Task} class has only one UDF, $t$.{\em compute}({\em frontier}), where the input {\em frontier} keeps the set of vertices requested by $t$ in its previous iteration. Each element of {\em frontier} is actually a pointer to a vertex object in $T_{local}$ or $T_{cache}$. Of course, users may also access $t$'s subgraph and context object in {\em compute}({\em frontier}).

UDF {\em compute}({\em frontier}) specifies how a task computes for one iteration. If $t$.{\em compute}(.) returns {\em true}, $t$ needs to be processed by more iterations; otherwise, $t$'s computation is finished after the current iteration. In \oursys, when $t$ is fetched from the task queue for processing, $t$.{\em compute}(.) is executed repeatedly until either $t$ is complete, or $t$ needs a non-local vertex $v$ that is not cached in $T_{cache}$, in which case $t$ is added to the task queue waiting for all $t$'s requested vertices to be pulled. When a task is completed or queued to disk, \oursys automatically garbage collects the memory space of the task to make room for the processing of other tasks.

Inside $t$.{\em compute}(.), a user may access and update $g$ and {\em context}, and call {\em pull}($u$) to request vertex $u$ for use in $t$'s next iteration. Here, $u$ is usually in the adjacency list of a previously pulled vertex, and {\em pull}($u$) expands the frontier of $g$. To improve network utilization, $g$ is usually expanded in a breadth-first manner, so that each call of {\em compute}(.) generates pull-requests for all relevant vertices adjacent to $g$'s growing frontier. A user may also call {\em add\_task}({\em task}) in $t$.{\em compute}(.) to add a newly-created task to the task queue.

\vspace{1mm}

\noindent{\bf The Worker Class.} Each object of the {\em Worker} class corresponds to a worker that processes its assigned tasks in serial. Figure~\ref{api_simple} shows the key functions of the {\em Worker} class, including two important UDFs.

UDF {\em seedTask\_gene}($v$) specifies how to create tasks according to a seed vertex $v\in T_{local}$. A worker of \oursys starts by calling {\em seedTask\_gene}($v$) on every $v\in T_{local}$, to generate seed tasks and to add them to the disk-based task queue. Inside {\em seedTask\_gene}($v$), users may examine the adjacency list of $v$, create tasks accordingly (and may let each task pull neighbors of $v$), and add these tasks to the task queue by calling {\em add\_task(.)}.

UDF {\em respond}($v$) is used to prune $\Gamma(v)$ before sending it back to requesting workers. By default, {\em respond}($v$) returns {\em NULL} and \oursys directly uses the vertex object of $v$ in $T_{local}$ to respond. Users may overload {\em respond}($v$) to return a newly created copy of $v$, with items in $\Gamma(v)$ properly pruned to save communication (e.g., $\Gamma_{gt}(v)$ for clique enumeration). In this case, \oursys will respond by sending the new object and then garbage-collect it.

The {\em worker} class also contains formatting UDFs, e.g., for users to define how to parse a line in the input file on HDFS into a vertex object in $T_{local}$, which will be used during graph loading.

To run a \oursys program, one may subclass {\em Worker} with all UDFs properly implemented, and then call {\em run}({\em config\_info}) to start the job, where {\em config\_info} contains job configuration parameters such as the HDFS file path of the input graph.

\vspace{1mm}

\noindent{\bf The Aggregator Class.} The {\em Worker} class optionally admits a second template argument $<${\em aggT}$>$, which needs to be specified if aggregator is used to collect some statistics such as triangle count or maximum clique size. Each task can aggregate a value to its worker's local aggregator when it finishes. These locally aggregated values can either be globally aggregated at last when all workers finish computing their tasks (which is the default setting), or be periodically synchronized (e.g., every 10 seconds) to make the globally aggregated value available to all workers (and thus all tasks) timely for use (e.g., in {\em compute}(.) to prune search space). In the latter case, users need to provide a frequency parameter.

\section{Applications}\label{sec:app}
We consider two categories of applications: (1)~finding dense subgraph structures such as triangles, cliques and quasi-cliques, which is useful in social network analysis and community detection; (2)~graph matching, which is useful in applications such as querying semantic networks and pattern recognition.

For simplicity, we only consider top-level task decomposition, i.e., we grow each vertex $v_i\in V$ into {\em exactly one} decomposed subgraph $G_i$, and every qualified subgraph will be found in {\em exactly one} decomposed subgraph.

\vspace{1mm}

\noindent{\bf Triangle Counting.} Assume that for any vertex $v$, neighbors in $\Gamma(v)$ are already sorted in increasing order of vertex ID (e.g., during graph loading). We also denote the largest (i.e., last) vertex in $\Gamma(v)$ by $\max_\Gamma(v)$.

We want each triangle $\triangle v_1v_2v_3$ (w.l.o.g., $v_1<v_2<v_3$) to be counted exactly once, i.e., in $v_1$'s decomposed subgraph. We let $v_1$ count $\triangle v_1v_2v_3$ by checking whether $v_3\in\Gamma(v_2)$. Since $v_1$ only examines $\Gamma(v_2)$ for every neighbor $v_2$ with $v_1<v_2<v_3$, $v_1$ only needs to pull the adjacency list of every neighbor in $(\Gamma_{gt}(v_1)-\{\max_\Gamma(v_1)\})$. Also, since $\Gamma(v_2)$ is only checked against $v_3>v_2$, $v_2$ only needs to respond $\Gamma_{gt}(v_2)$ to $v_1$.

According to the above discussion, among the UDFs of {\em Worker}, {\em respond}($v_2$) creates a copy of $v_2$ with adjacency list $\Gamma_{gt}(v_2)$ for responding; if $|\Gamma_{gt}(v_1)|\geq 2$, {\em seedTask\_gene}($v_1$) creates a task $t$ for $v_1$ and let $t$ pull every vertex in $(\Gamma_{gt}(v_1)-\{\max_\Gamma(v_1)\})$. The context of $t$ keeps $\max_\Gamma(v_1)$ and a triangle counter {\em count} (initialized as 0).

In $t$.{\em compute}({\em frontier}), {\em frontier} contains all the pulled vertices (i.e., $v_2$) in increasing order of their IDs as they were requested in {\em seedTask\_gene}($v_1$). We check every $v_2\in$ {\em frontier} as follows. For each $v_2$, we loop through all vertices $v_3>v_2$ in $\Gamma_{gt}(v_1)$ ($\Gamma_{gt}(v_1)$ is obtained by appending $\max_\Gamma(v_1)$ in $t$'s context to {\em frontier}), and increment $t$'s counter if $v_3\in\Gamma(v_2)$. Finally, {\em compute}(.) returns {\em false} since we have checked all $(v_2, v_3)$ pairs and the task is finished.

Whenever a task $t$ is finished, its counter (in $t$'s context) is added to the locally aggregated value, and when all workers finish computation, these local counts are sent to the master to get the total triangle count.

\vspace{1mm}

\noindent{\bf Maximum Clique.} We adapt the serial backtracking algorithm of~\cite{clique03} to \oursys. The original algorithm maintains the size of the maximum clique currently found, denoted by $|Q_{max}|$, to prune the search space.

To allow timely pruning, each worker in our \oursys program maintains $|Q_{max}|$ and keeps it relatively up to date by periodic aggregator synchronization, so that if a worker discovers a larger clique and updates $|Q_{max}|$, the value can be synchronized to other workers timely to improve their pruning effectiveness. In {\em seedTask\_gene}($v_i$), we create a task $t$ whose graph $g$ contains $v_i$, and we let $t$ pull all vertices in $\Gamma_{gt}(v_i)$. Then in $t$.{\em compute}({\em frontier}), we collect vertices in {\em frontier} (i.e., $\Gamma_{gt}(v_i)$), add them to $g$ but filter those adjacency list items that are not in $\{v_i\}\cup\Gamma_{gt}(v_i)$, to form the decomposed subgraph $G_i$, and then run the algorithm of~\cite{clique03} on $G_i$.

This solution can be easily extended to find quasi-cliques, where in a quasi-clique, every vertex is adjacent to at least $\gamma$ ($\geq0.5$) fraction of other vertices. In such a quasi-clique, two vertices are at most 2 hops away~\cite{quasiclique}. The \oursys algorithm is similar to that for finding maximum clique, except that (1)~for each local seeding vertex $v_i$, {\em compute}({\em frontier}) runs for 2 iterations to pull vertices (larger than $v_i$) within 2 hops of $v_i$; (2)~{\em compute}({\em frontier}) then constructs $G_i$ as the 2-hop ego-network of $v_i$ and runs the quasi-clique algorithm of~\cite{quasiclique} on $G_i$ to compute the quasi-cliques.

\vspace{1mm}

\begin{figure}[t]
    \centering
    \includegraphics[width=0.65\columnwidth]{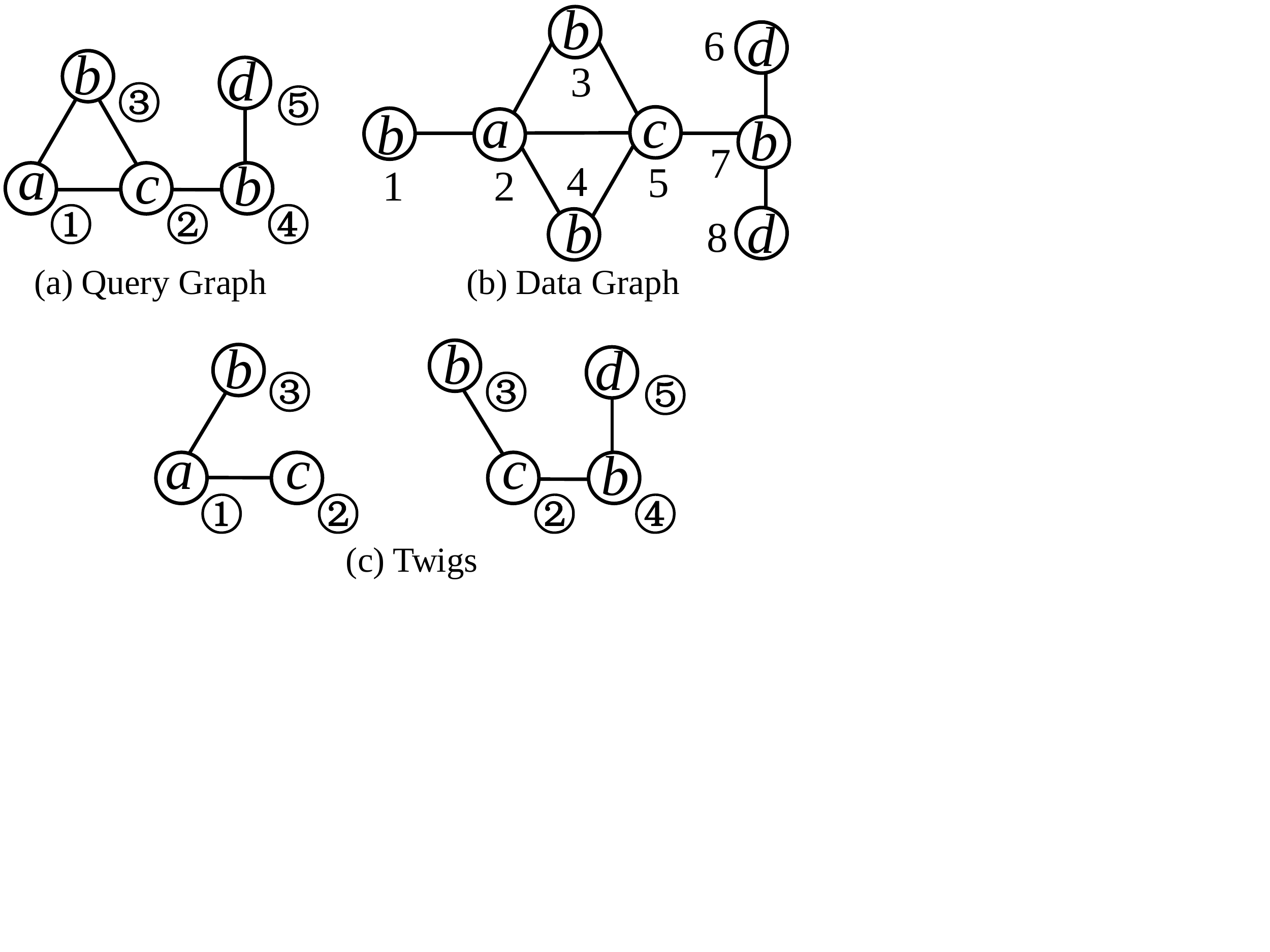}
    \caption{An Example of Graph Matching}\label{gmatch}
    \vspace{-4mm}
\end{figure}

\noindent{\bf Graph Matching.} Graph matching finds all subgraph instances in a data graph that match the query graph. Consider the problem of finding all occurrences of the query graph pattern given by Figure~\ref{gmatch}(a) in the data graph shown in Figure~\ref{gmatch}(b). In this example, each vertex in the query graph (and the data graph) has a unique integer ID and a label. We define $k_1k_2k_3k_4k_5$ as a mapping where vertex with ID $k_i$ in the data graph is mapped to vertex $\textcircled{i}$ in the query graph. A mapping is a matching if vertex $k_i$ and vertex $\textcircled{i}$ have the same label (for any $i$), and for any edge $(\textcircled{i}, \textcircled{j})$ in the query graph, the corresponding edge $(k_i, k_j)$ exists in the data graph. For example, 25478 is a matching, while 25178 is not since the data graph does not have edge $(1, 5)$ that corresponds to $(\textcircled{3}, \textcircled{2})$.

Existing works on distributed graph matching combine vertex-centric graph exploration with subgraph join. Note that when a query graph contains cycles, vertex-centric graph exploration alone is not sufficient. For example, in Figure~\ref{gmatch}(b), suppose that we perform vertex-centric exploration on the data graph along query graph path $\textcircled{3}$-$\textcircled{1}$-$\textcircled{2}$, we will explore from Vertex~1 (or~4) to~2 and then to~5 simply according to neighbors' labels. Then, we need to check all $b$-labeled neighbors of Vertex~5 to find Vertex~1 (or~4), which is essentially an equi-join on the ID of $k_3$ rather than a simple label-based exploration. \cite{match1} and~\cite{match2} first decompose a query graph into small acyclic subgraphs called twigs (see Figure~\ref{gmatch}(c) for an example), and then use graph exploration to find subgraph instances that match those twigs, and join twigs on joint vertices (e.g., $k_2$ and $k_3$ for Figure~\ref{gmatch}(c)) to obtain the subgraphs that match the query graph.

Our algorithm avoids materializing matched subgraphs and performing distributed subgraph join as required by existing solutions. Instead, we pull required vertices to construct each decomposed subgraph $G_i$, and then simply enumerate the matched subgraph instances in the decomposed graph using backtracking without generating any communication.

We illustrate how to write a \oursys program for the query graph of Figure~\ref{gmatch}(a). Assume that each adjacency list item contains vertex label\footnote{If this is not the case, one may use the Pregel algorithm of~\cite{da_www} for attribute broadcast to preprocess the graph data in linear cost.}. We start the matching from vertex \textcircled{1} with label ``$a$'', and grow $G_i$ from each vertex $v_i$ in the data graph with label ``$a$''. Note that every matched subgraph instance will be found since it must contain an $a$-labeled vertex $v_i$, and it will only be found in $v_i$'s decomposed subgraph $G_i$.

We now present our algorithm, which can be safely skipped if you are not interested in reading the details.

\vspace{1mm}

\noindent{\bf The Algorithm:} in {\em seedTask\_gene}($v$), we only create a task $t$ for $v$ if $v$'s label is ``$a$'', and $\Gamma(v)$ contains neighbors with both labels ``$b$'' and ``$c$''. If this is the case, we add vertex $v$ to $g$, and pull all vertices in $\Gamma(v)$ with labels ``$b$'' and ``$c$''.

Then, in iteration~1 of $t$.{\em compute}({\em frontier}), we split {\em frontier} into two vertex sets: $V_b$ (resp.\ $V_c$) consists of vertices with label ``$b$'' (resp.\ ``$c$''). However, while a vertex in $V_c$ definitely matches vertex \textcircled{2} in Figure~\ref{gmatch}(a), a vertex in $V_b$ may match either Vertex~\textcircled{3} or Vertex~\textcircled{4}. For each vertex $v_c\in V_c$, we split all vertices in $\Gamma(v_c)$ with label ``$b$'' into two sets: $U_1$ consisting of those vertices that are also in $V_b$ (i.e., they can match Vertex~\textcircled{3} or Vertex~\textcircled{4}), and $U_2$ consisting of the rest (i.e., they can only match Vertex~\textcircled{4} since they are not neighbors of $v_c$). We prune $v_c$, (i)~if $U_1=\emptyset$ since $v_c$ does not have a neighbor matching Vertex~\textcircled{3}, or (ii)~if $|U_1|=1$ and $U_2=\emptyset$, since $v_c$ does not have two neighbors with label ``$b$''. Otherwise, (iii)~if $|U_1|=1$ and $U_2\neq\emptyset$, then the vertex in $U_1$ has to match Vertex~\textcircled{3}, and the vertex matching Vertex~\textcircled{4} has to be from $U_2$, and thus we pull all vertices of $U_2$; while (iv)~if $|U_1|>1$, the vertex matching Vertex~\textcircled{4} can be from either $U_1$ or $U_2$, and thus we pull all vertices from both $U_1$ and $U_2$. Let the only vertex (with label ``$a$'') currently in $g$ be $v_a$, then in both Cases~(iii) and~(iv), we add $v_c$ and edge $(v_a, v_c)$ to $g$, and for each vertex $v_b\in U_1$ (i.e., $v_b$ can match Vertex~\textcircled{3}), we add $v_b$ and edge $(v_a, v_b)$ to $g$.

Then in iteration~2 of $t$.{\em compute}({\em frontier}), {\em frontier} contains all pulled vertices with label ``$b$'' that can match Vertex~\textcircled{4}. Let the set of all vertices with label ``$c$'' in $g$ (i.e., matching Vertex~\textcircled{2}) be $V_c$. Then, for each vertex $v_b\in$ {\em frontier}, we denote the set of all vertices of $\Gamma(v_b)$ with label ``$d$'' (i.e., matching Vertex~\textcircled{5}) be $V_d$; if $V_d\neq\emptyset$, (1)~we add $v_b$ to $g$, (2)~for every vertex $v_c\in V_c\cap\Gamma(v_b)$, we add edge $(v_c, v_b)$ (i.e., matching $(\textcircled{2}, \textcircled{4})$) to $g$, (3)~for every vertex $v_d\in V_d$, we add $v_d$ and edge $(v_b, v_d)$ to $g$. Finally, we run a backtracking algorithm on $g$ to enumerate all subgraphs that match the query graph.

Lastly, we can let UDF {\em respond}($v$) return a copy of $v$ by pruning items in $\Gamma(v)$ whose labels do not fall into \{a, b, c, d\} to save communication. \qed

\vspace{1mm}

We remark that only top-level subgraphs decomposed by Vertex~\textcircled{1} in the query vertex have been considered. If a resulting decomposed subgraph $G_i$ is still too big, one may continue to decompose $G_i$ by looking at one more vertex in the query graph (given that Vertex~\textcircled{1} is already matched to $v_i$).

\section{System Implementation}\label{sec:system}

\begin{figure}[t]
    \centering
    \includegraphics[width=0.8\columnwidth]{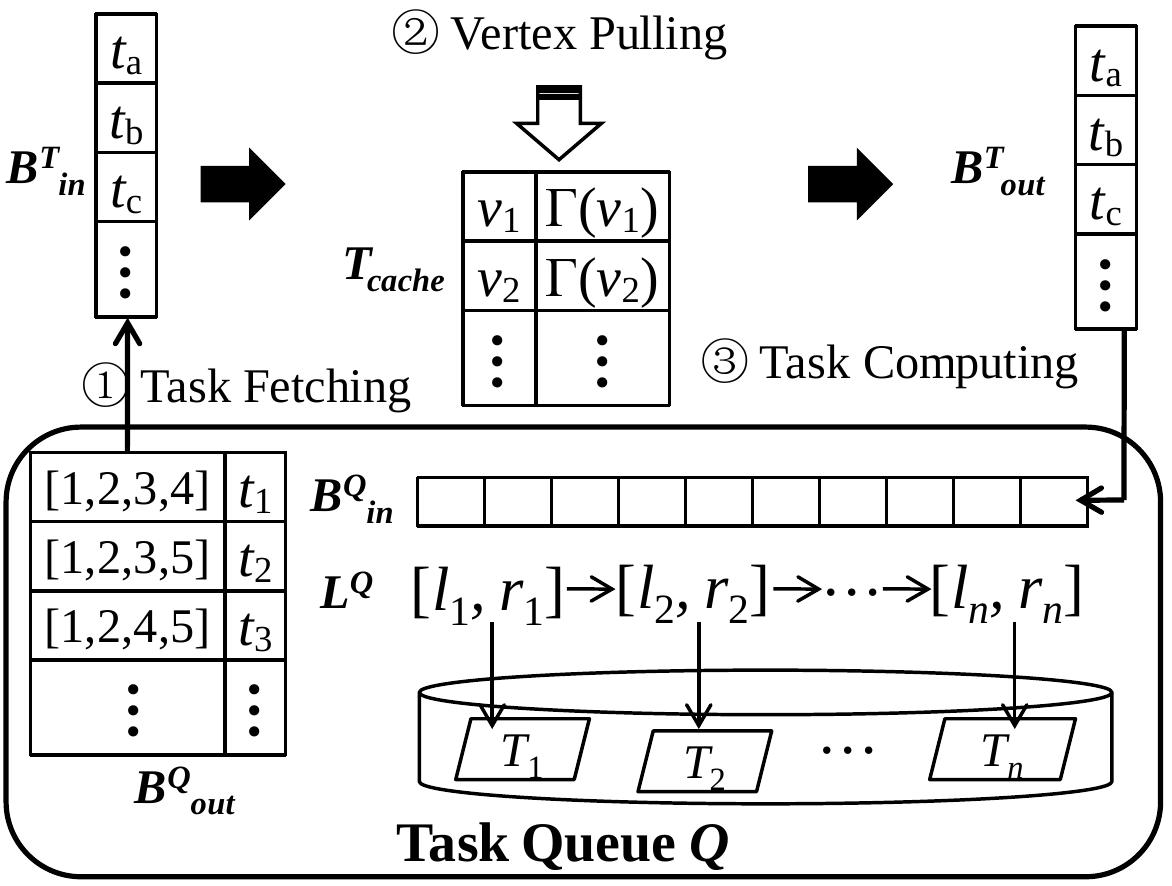}
    \caption{Computation Framework of a Worker}\label{frame}
    \vspace{-4mm}
\end{figure}

\noindent{\bf Task Queue \& Task Buffers.} Since tasks contain subgraphs that may overlap with each other, it is impractical to keep all tasks in memory. Thus, each worker of \oursys maintains a disk-based queue $Q$ to keep those tasks waiting to be processed. Figure~\ref{frame} shows the procedure of task computation on a worker of \oursys. Specifically, tasks are fetched from the task queue $Q$ one at a time and added to a task buffer $B^T_{in}$ for batch processing, while the processed tasks (and those newly-created by {\em add\_task}(.)) are appended to buffer $B^T_{out}$ and then merged to $Q$ in batches.

One baseline approach to organize $Q$ is to treat it as a local-disk stream of tasks, which allows tasks to be sequentially read from (and appended to) $Q$. To support two-sided streaming, we organize tasks in $Q$ with files each containing $C$ tasks, where $C$ is a user-defined parameter (100 by default) to amortize the random IO cost of reading (and writing) a task file. We call this queue organization as {\bf\em stream-queue}, which does not consider whether tasks fetched into $B^T_{in}$ share common vertices to pull. To increase the probability that tasks in $B^T_{in}$ share common vertices to pull, we designed another queue organization called {\bf\em LSH-queue} based on min-hashing.

Specifically, assume that a task $t$ has called {\em compute}(.), and let us denote the set of vertices that $t$ needs to pull from remote machines by $\mathbf{P(t)}$. Before adding $t$ to $Q$, we append a key $k(t)$ to $t$, which consists of a sequence of $\ell$ ($=4$ by default) MinHash signatures~\cite{minhash} of $P(t)$. Due to the locality sensitivity of min-hashing, for two tasks $t_1$ and $t_2$, the more similar $k(t_1)$ and $k(t_2)$ are, the more likely $P(t_1)$ and $P(t_2)$ overlap~\cite{minhash}. Therefore, we keep tasks in $Q$ ordered by their keys (in alphabetic order), so that tasks fetched tasks from the head of $Q$ have similar keys (e.g., see $B^Q_{out}$ in Figure~\ref{frame}) and are likely to share more common vertices to pull. Reducing redundancy by ordering data according to min-hashing keys has been shown to be effective by previous works~\cite{nscale,eagr}.

To avoid random disk IOs, we organize $Q$ as depicted in Figure~\ref{frame}. We maintain an in-memory buffer $B^Q_{in}$ to receive incoming tasks (from $B^T_{out}$), and an in-memory buffer $B^Q_{out}$ to buffer ordered tasks to be fetched. The waiting tasks on local disk are grouped into files, where each file contains $[C/2, C]$ tasks ordered by their keys and $C$ is a user-defined parameter. The ranges of keys in different files are disjoint (except at boundaries where keys may be equal), and all task files are linked in the order of key ranges by an in-memory doubly-linked list $L^Q$. Each element in $L^Q$ points to a task file and records its key range. Here, $L^Q$ is like the leaf level of a B$^+$-tree, but it is small enough to be memory-resident since only metadata of files are kept. When $B^Q_{in}$ overflows, we merge all its tasks to the list of task files efficiently by utilizing $L^Q$ while guaranteeing that each file still contains $[C/2, C]$ tasks after merging, using a B$^+$-tree style algorithm.

A computing thread fetches tasks one by one from $B^Q_{out}$ for computation, and when $B^Q_{out}$ becomes empty, we load to $B^Q_{out}$ those tasks in the first file of $L^Q$ if $L^Q$ is not empty; otherwise, we fill $B^Q_{out}$ with tasks obtained from the head of $B^Q_{in}$. Since tasks in a file are clustered by their keys, tasks in $B^Q_{out}$ tend to share common vertices to pull.

\vspace{1mm}

\noindent{\bf Task Computation.} A worker of \oursys processes its tasks in rounds, where each round consists of three steps, (1)~\emph{task fetching}, (2)~\emph{vertex pulling}, and (3)~\emph{task computing}. Figure~\ref{frame} illustrates these three steps.

The first step fetches tasks from $Q$ into $B^T_{in}$ until either (i)~$B^T_{in}$ becomes full, or (ii)~there is no more room in $T_{cache}$ to accommodate more vertices to pull. The second step then pulls all requested vertices that are not already hit in $T_{cache}$. Note that the pulling frequency is influenced by the capacity of $B^T_{in}$ and $T_{cache}$. Now, for every task $t\in B^T_{in}$, its requested vertices in {\em frontier} are either in $T_{local}$ or $T_{cache}$, and thus we start the third step to process the tasks in $B^T_{in}$. We compute each task $t$ iteratively until either $t$ is complete, or there exists a newly-requested vertex that is neither in $T_{local}$ nor in $T_{cache}$. In the latter case, we compute $t$'s key using $P(t)$, and then add $t$ to $B^T_{out}$. If new tasks are created by $t$, they are also added to $B^T_{out}$. Whenever $B^T_{out}$ is full, it merges its tasks to $Q$.

Since a machine runs multiple workers, and the independence of their execution allows computation and communication to overlap.

\vspace{1mm}

\noindent{\bf Other Issues.} Real graphs may contain some high-degree vertex $v$, and the task seeded from $v$ may have $|P(t)|$ larger than the capacity of $T_{cache}$. To allow such a task to proceed, we treat $t$ as a singleton task batch to perform vertex pulling, by temporarily increase the capacity of $T_{cache}$. After $t.${\em compute}(.) returns, we recover the original capacity of $T_{cache}$ by evicting overflowed vertices, before starting the next round.

A worker initially seeds the tasks from all vertices in $T_{local}$ into $Q$, to maximize the opportunity of finding tasks that share common vertices to pull. The seeded tasks are merge-sorted by their min-hashing key to efficiently create the file list of $Q$.

There exists some work that uses heuristics to estimate the computation cost of a task $t$ from its decomposed subgraph~\cite{clique13}, and an online regressor may also be trained to improve the cost estimation after each task is finished. Since estimating the cost of $t$ from a partially grown subgraph is difficult, we only estimate $t$'s cost when its decomposed subgraph is fully constructed. To allow task prefetching, each worker buffers a small set of tasks with estimated costs, and if all other tasks are exhausted, the work requests tasks from a coordinating master while continuing to process the buffered tasks. The master collects task summary from all workers to decide which tasks to be redistributed when some worker requests more tasks. We remark that task stealing strategies are still under development and are thus not reported in this paper.

\section{Experiments}\label{sec:results}
We evaluate the performance of \oursys, which is implemented in C++ and communicates with HDFS (Hadoop 2.6.0) using libhdfs. All our experiments were run on a cluster of 15 machines, each with 12 cores (two Intel Xeon E5-2620 CPUs) and 48GB RAM. The connectivity between any pair of nodes in the cluster is 1Gbps. All system and application codes are released in \oursys's website~\footnote{http://yanda.cis.uab.edu/gthinker/}. We report end-to-end processing time, from graph loading to when the slowest worker finishes its processing, for all the following experiments.

\begin{table}[t]
\caption{Graph Datasets (M = 1,000,000)}\label{data}
\centering
\includegraphics[width=0.96\columnwidth]{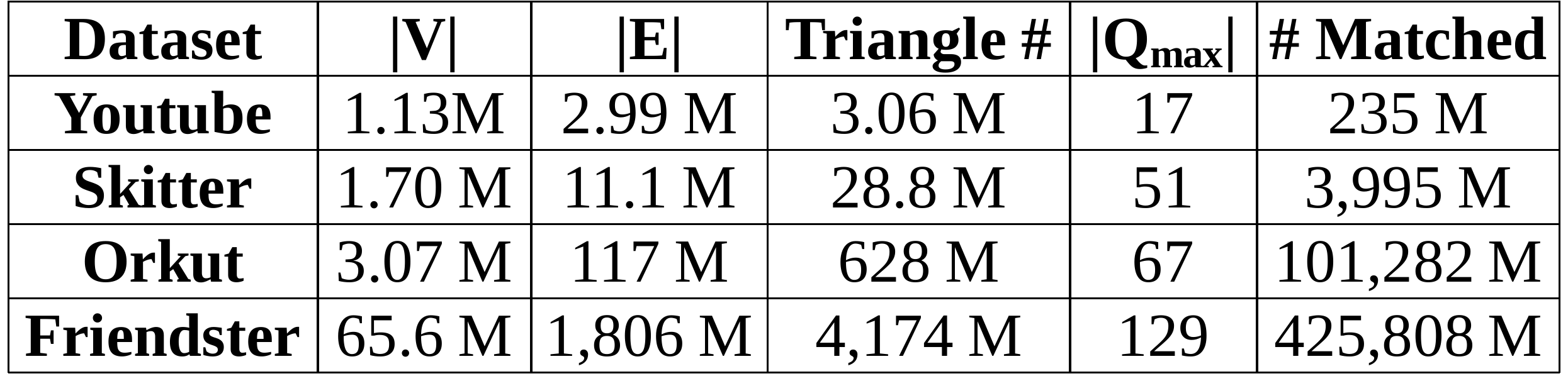}
\vspace{-4mm}
\end{table}

Table~\ref{data} shows the four real-world graph datasets used in our experiments. We chose these graphs to be undirected since our applications described in Section~\ref{sec:app} are for undirected graphs, while \oursys can also handle directed graphs. These graphs are also chosen to have different sizes: {\em Youtube}\footnote{https://snap.stanford.edu/data/com-Youtube.html}, {\em Skitter}\footnote{http://konect.uni-koblenz.de/networks/as-skitter}, {\em Orkut}\footnote{http://konect.uni-koblenz.de/networks/orkut-links}, and {\em Friendster}\footnote{http://snap.stanford.edu/data/com-Friendster.html} have 2.99 M, 11.1 M, 117 M and 1,806 M undirected edges, respectively.

We ran the algorithms described in Section~\ref{sec:app}, and list the triangle count, maximum clique size (denoted by $|Q_{max}|$), and number of matched subgraph instances in Table~\ref{data}. For graph matching, we used the query graph of Figure~\ref{gmatch}(a) and randomly generated a label for each vertex in the data graph among $\{a,b,c,d,e,f,g\}$ (following a uniform distribution). We can see that the job is highly computation-intensive; e.g., the number of matched subgraphs is in the order of $10^{11}$ for {\em Orkut} and {\em Friendster}.

Recall from Table~\ref{frame} that each worker in \oursys maintains four task buffers, $B^T_{in}$, $B^T_{out}$, $B^Q_{in}$ and $B^Q_{out}$. We set the capacity of $B^T_{in}$, $B^T_{out}$ and $B^Q_{in}$ to be the same, which is the maximum number of tasks that are processed in each round. We call this capacity as {\bf\em buffer capacity}. We also set the capacity of $B^Q_{out}$ to be the {\bf\em file capacity} $C$, since it loads a file of tasks from disk each time.

Unless otherwise stated, the default setting is as follows. Each machine runs 8 workers (i.e., 120 workers in total). The buffer capacity is set as 1000 tasks, and the file capacity $C$ is set to be 100 tasks. Moreover, we set $T_{cache}$ to accommodate up to 1 M non-local vertices.

\vspace{1mm}

\begin{table}[t]
\caption{Comparison with Serial Algorithms}\label{serial}
\centering
\includegraphics[width=\columnwidth]{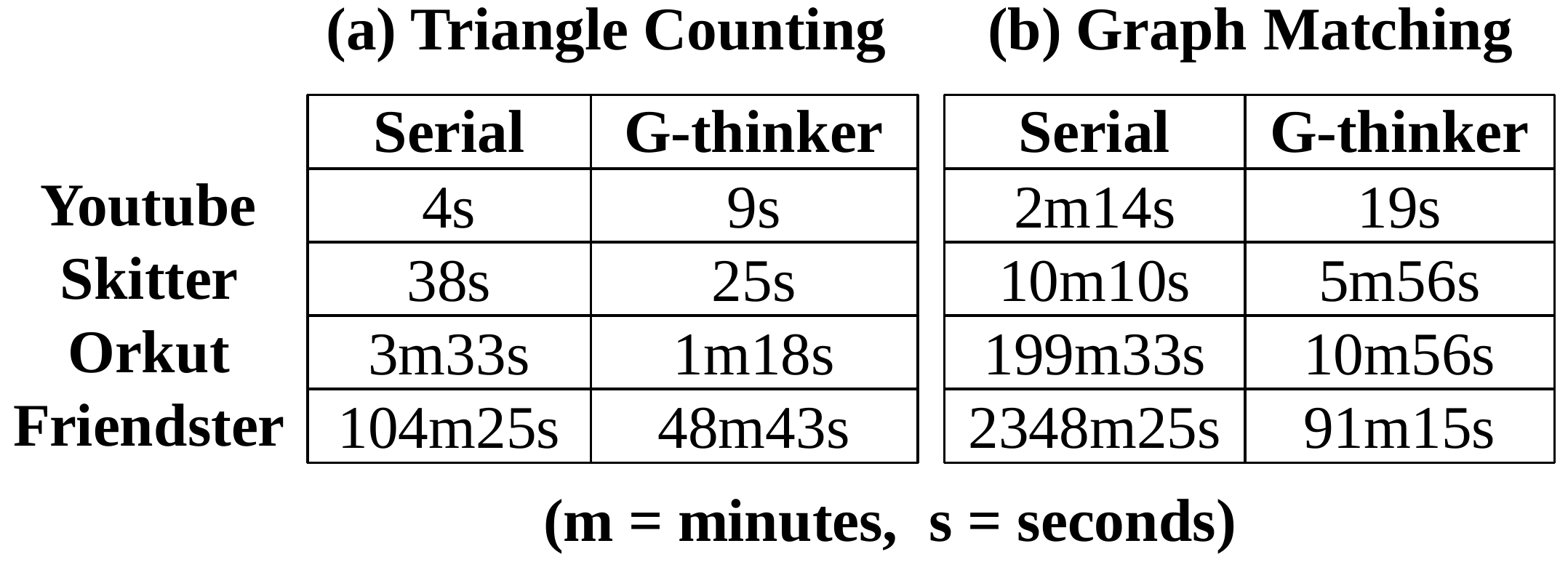}
\vspace{-6mm}
\caption{System Comparison (Triangle Counting)}\label{compare}
\includegraphics[width=0.86\columnwidth]{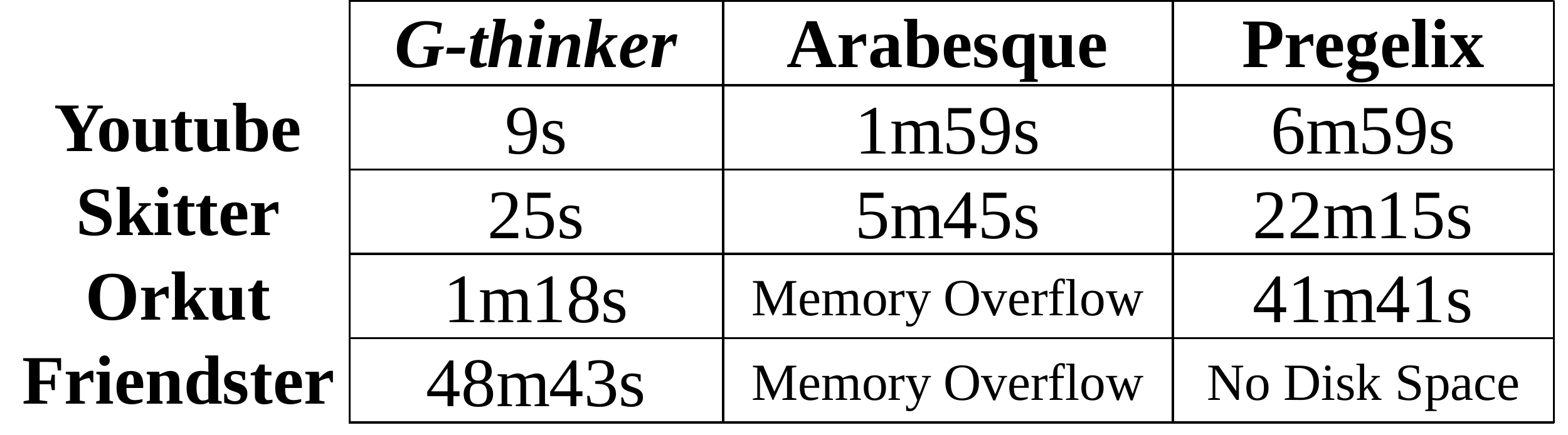}
\vspace{-4mm}
\end{table}

\begin{table*}[htbp]
\begin{minipage}[t]{0.6\linewidth}
\caption{Scalability}\label{scale}
\centering
\includegraphics[width=\columnwidth]{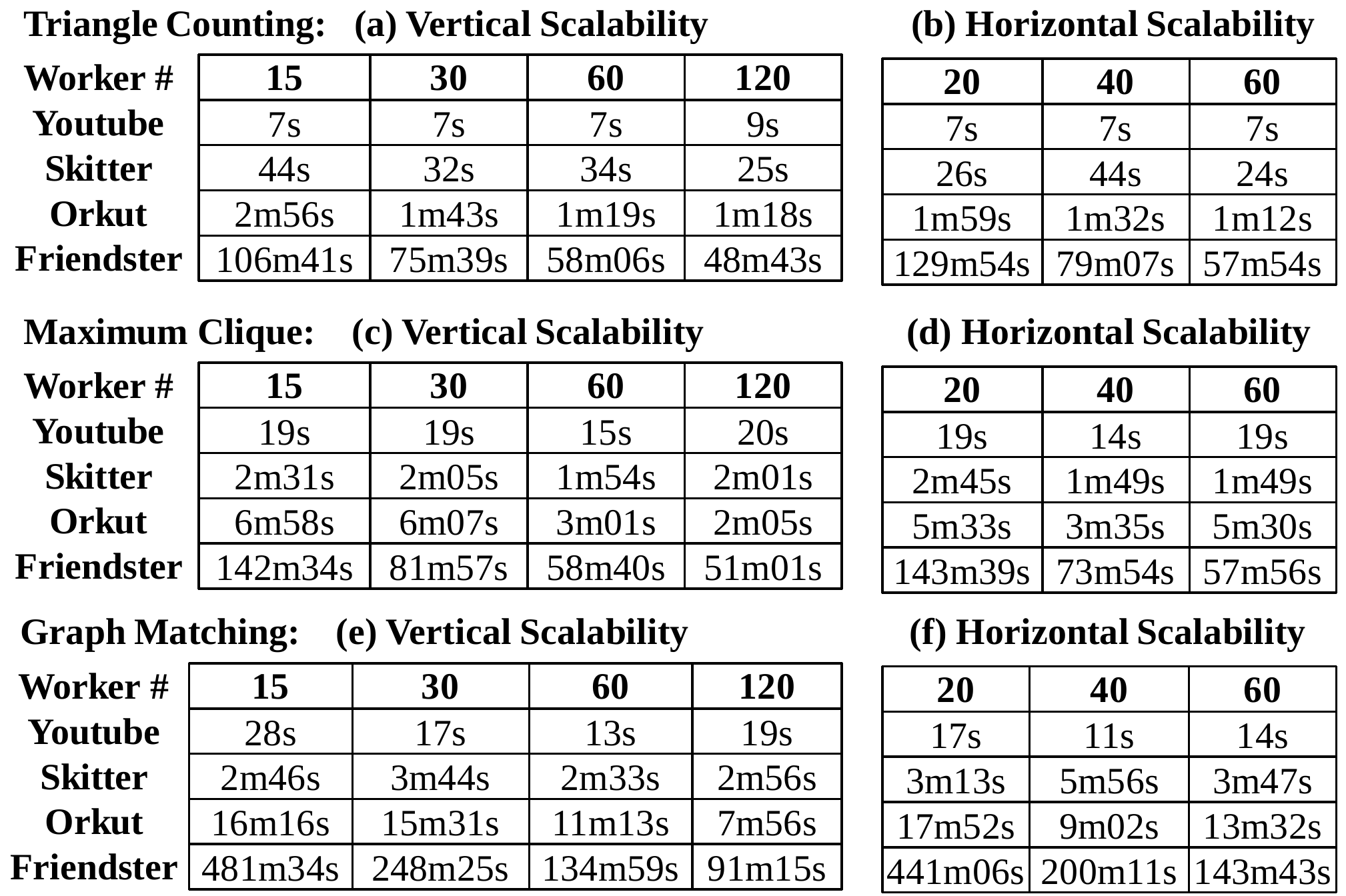}
\vspace{-4mm}
\end{minipage}
\hfill
\begin{minipage}[t]{0.38\linewidth}
\caption{Effect of System Parameters}\label{param}
\centering
\includegraphics[width=\columnwidth]{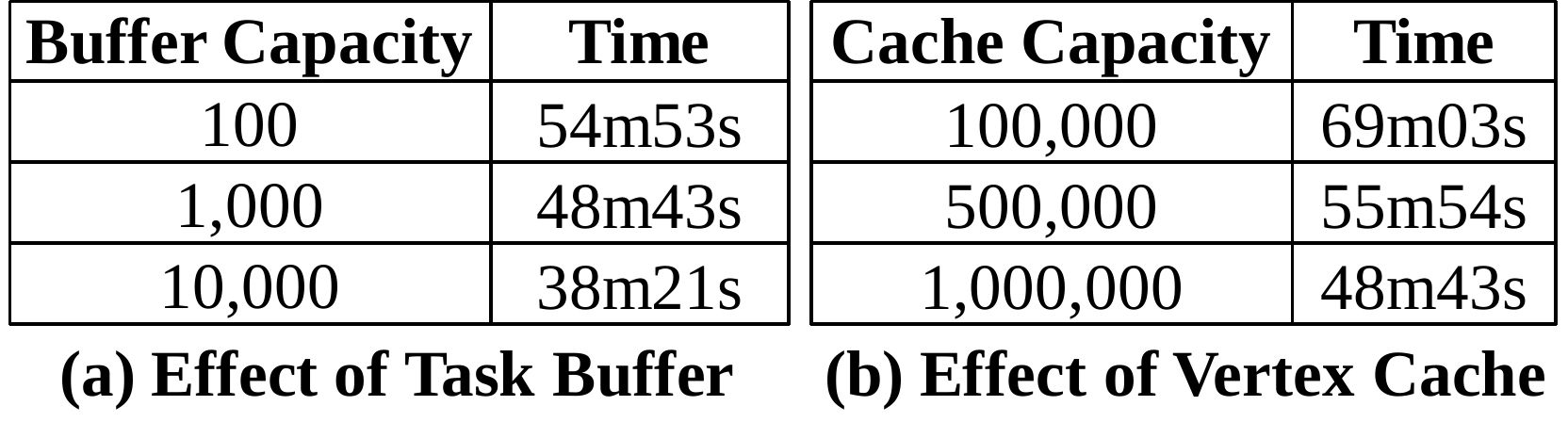}
\vspace{-6mm}
\caption{\# of Random IO by LSH-Queue}\label{io}
\centering
\includegraphics[width=\columnwidth]{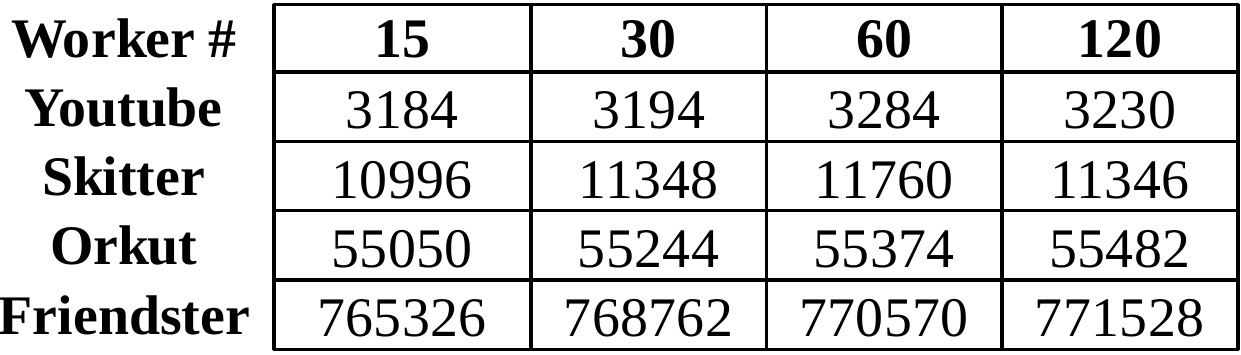}
\vspace{-6mm}
\caption{Results with Stream-Queue}\label{stream_queue}
\centering
\includegraphics[width=\columnwidth]{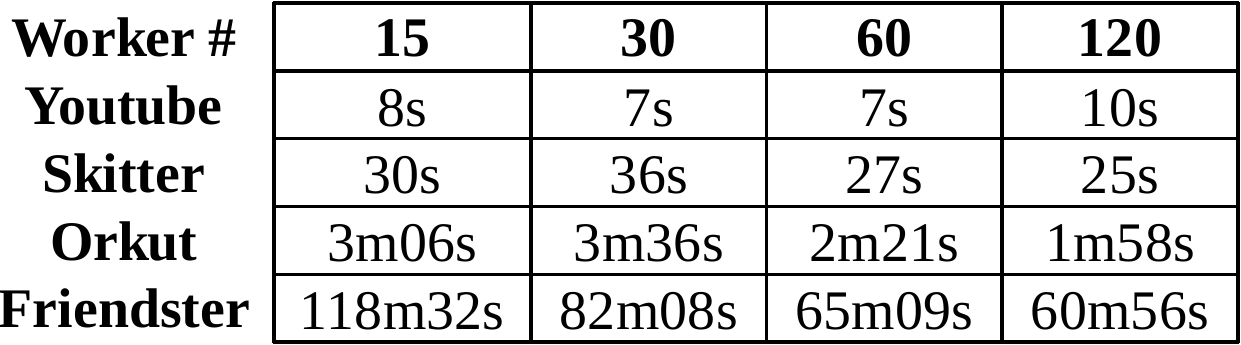}
\end{minipage}
\vspace{-4mm}
\end{table*}

\noindent{\bf Comparison with Serial Algorithms.} We first compare the performance of the serial algorithms for subgraph finding with their distributed \oursys counterparts. Since serial algorithms need to hold an entire input graph in memory, we run them in a high-end machine with 1TB DDR3 RAM and 2.2GHz CPUs. Table~\ref{serial} reports the comparison results for triangle counting (with relatively low computation intensity) and graph matching (which is highly computation-intensive). We see that \oursys is orders of magnitude faster than the serial algorithm for graph matching, but only several times faster for triangle counting. This is because the light computation workload of triangle counting cannot offset the communication cost of vertex-pulling.

\vspace{1mm}

\noindent{\bf Comparison with Other Systems.} We also compare \oursys with Arabesque~\cite{arab} and Pregelix (version 0.2.12)~\cite{pregelix}. Both Arabesque and Pregelix have already implemented triangle counting and maximal clique enumeration, and we used these programs directly for comparison. Unfortunately, neither Arabesque nor Pregelix could successfully finish maximal clique enumeration on even the smallest graph {\em Youtube} in our cluster. Arabesque failed after running for 1.5 hours due to memory overflow, while Pregelix reported a frame size error since the 1-ego network of some vertex cannot fit in a frame as required by Pregelix. In contrast, \oursys found the maximum clique (of size 17) in {\em Youtube} in just around 20 seconds.

Table~\ref{compare} reports the results for triangle counting. We can see that \oursys is orders of magnitude faster than both Arabesque and Pregelix, while memory-based Arabesque is a few times faster than disk-based Pregelix. However, Arabesque failed to process {\em Orkut} and {\em Friendster} due to insufficient memory space. Pregelix failed to process {\em Friendster} because it used up the disk space, probably due to its space-consuming B-tree structure for storing vertex data.

Finally, although NScale~\cite{nscale} is not public, \cite{nscale} reported that it takes 1986 seconds to count the triangles of {\em Orkut} (not including the expensive subgraph construction \& packing) on a 16-node cluster, while \oursys takes only 78 seconds on our 15-node cluster.

\vspace{1mm}

\noindent{\bf Scalability.} We tested the vertical scalability of \oursys by running various applications with all 15 machines, where each machine runs 1, 2, 4 and 8 workers, respectively. We also tested the horizontal scalability of \oursys by running various applications with 5, 10 and 15 machines, where each machine runs 4 workers. The results are reported in Table~\ref{scale}, where we can see that \oursys scales well with the number of workers per machine, and the total number of machines, especially for highly computation-intensive problems like graph matching.

\vspace{1mm}

\noindent{\bf Effect of System Parameters.} We conducted extensive experiments to study the impact of various parameters on system performance, and find that the performance is mainly sensitive to {\em buffer capacity} and the capacity of $T_{cache}$; the impact of other parameters such as the file capacity $C$ is minor. Due to space limit, we only report the results for triangle counting over {\em Friendster} here. Table~\ref{param}(a) shows the results when we change the buffer capacity while keeping all other parameters as default. We can see that the runtime decreases as the capacity of task buffers increase, but increasing the capacity beyond $10^4$ does not lead to much improvement. Table~\ref{param}(b) shows the results when we change the capacity of $T_{cache}$ while keeping all other parameters as default. We can see an obvious reduction in runtime as the capacity of vertex cache increases. Overall, the performance difference is not significantly influenced by the system parameters (e.g., less than doubled), and thus \oursys is expected to perform well even when the memory space is limited.

\vspace{1mm}

\noindent{\bf Stream-Queue v.s.\ LSH-Queue.} Due to space limit, we only report results for triangle counting; the results for maximum clique and graph matching are similar. Table~\ref{io} reports the total number of random disk reads and writes incurred by LSH-queue during the whole period of job execution, for the vertical scalability experiments we reported in Table~\ref{scale}(a). We can see that the number of random IO is small, which demonstrates that LSH-Queue exhibits near-sequential disk IO. We also repeated the experiments using stream-queue instead of LSH-queue, and Table~\ref{stream_queue} reports the results. Comparing Table~\ref{scale}(a) to Table~\ref{stream_queue}, we can see that LSH-Queue improves the performance of most jobs, especially those with heavy computation workload. For some jobs with relatively light workload, the overhead incurred by LSH-Queue (e.g., key computation and random IO) stands out and stream-queue is more efficient.

\vspace{-1mm}

\section{Conclusions and Future Work}\label{sec:conclude}
\vspace{-1mm}

We presented a new framework called \oursys for scalable subgraph finding, whose computation-intensive execution engine beats existing data-intensive systems by orders of magnitude, and scales to graphs with size two orders of magnitude larger given the same hardware resources. Future work of \oursys include designing effective task stealing strategy for load balancing, and acceleration through new hardware (e.g., SSD and GPU).

\bibliographystyle{abbrv}
\bibliography{ref_subg}

\end{document}